# Topological mechanochemistry of graphene

**Elena F. Sheka, Vera A. Popova, Nadezhda A. Popova**

Peoples' Friendship University of Russia

In view of a formal topology, two common terms, namely, connectivity and adjacency, determine the 'quality' of C-C bonds of $sp^2$ nanocarbons. The feature is the most sensitive point of the inherent topology of the species so that such external action as mechanical deformation should obviously change it and result in particular topological effects. The current paper describes the effects caused by uniaxial tension of a graphene molecule in due course of a mechanochemical reaction. Basing on the molecular theory of graphene, the effects are attributed to both mechanical loading and chemical modification of edge atoms of the molecule. The mechanical behavior is shown to be not only highly anisotropic with respect to the direction of the load application, but greatly dependent on the chemical modification of the molecule edge atoms thus revealing topological character of the graphene deformation.

## 1. Introduction

The modern topology in chemistry covers two large valleys, namely, formal, mathematical and empirical, chemical. The former is concerned with the description of molecular structures on the basis of finite topological spaces. The space shows itself as a mathematical image or instrument of theoretical study. A large collection of comprehensive reviews, related to a topological description of fullerenes from this viewpoint, has recently been published [1]. The second field covers vastly studied topochemical reactions. The space in this case is a physical reality defining the real place where the reactions occur. If the appearance of mathematical topology in chemistry can be counted off the publication of the Merrifield and Simmons monograph in 1989 [2], topochemical reactions have been studying from the nineteenth century (see [3] and references therein). The first stage of the study was completed in the late nineteenth-twenties [4] and then obtained a new pulse after appearing the Woodward and Hoffman monograph, devoted to the conservation of orbital symmetry, in 1970 [5]. Since then, topochemical reactions have become an inherent part of not only organic, but inorganic chemistry, as well. The readers, who are interested in this topic, are referred to a set of comprehensive reviews [3, 6-9], but a few. The current situation in this field can be seen by the example of a direct structural understanding of a topochemical solid state photopolymerization reaction [10].

Nowadays, we are witnessing the next pulse, stimulating investigations in the field, which should be attributed to the appearance of a new class of spatially extended molecular materials, such as $sp^2$ nanocarbons. Obviously, the main members of the class such as fullerenes, nanotubes, and numerous graphene-based species are absolutely different from the formal topology viewpoint. Thus, fullerenes exist in the form of a hollow sphere, ellipsoid, or tube consisting of differently packed benzenoid units. Carbon nanotubes present predominantly cylindrical packing of the units. In graphene, the benzenoid units form one-atom-thick planar honeycomb structure. If we address the common terms of the formal topology, namely, connectivity and adjacency, we have to intuitively accept their different amount in the above three species. In its turn, the connectivity and adjacency determine the 'quality' of the C-C bond structure of the species, thus, differentiating them by this mark. Since non-saturated C-C bonds are the main target for chemical reactions of any type, one must assume that identical reactions, involving the bonds, will occur differently for different members of the $sp^2$ nanocarbon family. Therefore, one may conclude that the spatially extended $sp^2$ nanocarbons present not only peculiarly structural chemicals, but the class of species for which the formal and empirical topology overlap. At the first time, results, presented in [11, 12] have revealed this tight interconnection in terms of molecular quantum theory. Not only fullerenes, but carbon nanotubes and graphene (their fragments) have been considered at the molecular level. The obtained results are related to the computational study of the intermolecular interaction between one of the above $sp^2$ nanocarbon molecules and one of the other addends, among which there are both $sp^2$ nanocarbons and monoatomic species. The intermolecular interaction lays the foundation of any reaction, so that its topological peculiarities may evidence a topochemical character of the reaction under study. However, since the 'quality' of C-C bonds is the most sensitive point of the inherent topology of $sp^2$ nanocarbons, external actions, say, mechanical deformation, on the bonds should obviously result in particular topological effects that accompany the relevant intramolecular reactions. The current paper is devoted to the discussion of such reactions that are presented by a mechanochemical one related to uniaxial tension of a graphene molecule.

## 2. Uniaxial tension of graphene as a mechanochemical reaction

Below we will consider a particular topological effect caused by the influence of both the loading direction and the graphene molecule edge termination on the inherited topology of the molecule. As turned out, the graphene deformation under external mechanical loading is extremely sensitive to the state of the edge atoms and makes it possible to disclose a topological nature of this sensitivity.

Oppositely to real physical experiments, when changing the object shape under loading is usually monitored, computational experiments deal with the total energy response to the object shape deformation that simulates either tension and contraction or bending, screwing, shift, and so forth. As for graphene, whose mechanical properties are amenable to experimental study with difficult, the computational experiments takes on great significance.

A lot of works are devoted to the calculation of mechanical properties of graphene due to which two approaches, namely, continuum and atomistic ones have been formulated. The continuum approach is based on the well developed theory of elasticity of continuous solid media applied to shells, plates, beams, rods, and trusses. The latter are structure elements used for the continuum description. When applying to graphene, its lattice structure is presented in terms of the above continuum structure elements and the main task of the calculation is the reformulation of the total energy of the studied atomic-molecular system subjected to changing in shape in terms of the continuum structure elements. This procedure involves actually the adaptation of the theory of elasticity of continuous media to nanosize objects which makes allowance for introducing macroscopic basic mechanical parameters such as Young's modulus (E), the Poisson ratio ($\nu$), the potential energy of the elastic deformation, etc into the description of mechanical properties of graphene. Since the energy of graphene is mainly calculated in the framework of quantum chemistry, which takes the object atom structure into account, the main problem of the continuum approach is a linkage between molecular configuration and continuum structure elements. Nanoscale continuum methods (see Refs. 13-17 and references therein), among which those based on the structural mechanics concept [18] are the most developed, have shown the best ability to simulate nanostructure materials. In view of this concept, graphene is a geometrical frame-like structure where the primary bonds between two nearest-neighboring atoms act like load-bearing beam members, whereas an individual atom acts as the joint of the related beams [19-22].

The basic concept of the atomistic approach consists in obtaining mechanical parameters of the object from results of the direct solutions of either Newton motion laws [22, 23] or Schrödinger equations [24, 25] under changing the object shape following a particular algorithm of simulation of the wished type of deformation. It should be necessary to issue a general comment concerning calculations based on the application of the DFT computational schemes. All the latter, except the recent one [26], were performed in the framework of restricted versions of the programs that do not take into account spins of the graphene odd electrons and thus ignore the correlation interaction between these electrons. The peculiarities of the graphene odd electron behavior are connected with a considerable enlarging of its C-C bonds, which, in its turn, causes a noticeable weakening of the odd electron interaction and thus requires taking into account these electrons correlation [27, 28].

In the case of atomistic approach, not energy itself, but forces applied to atoms become the main goal of calculations. These forces are input later into the relations of macroscopic linear theory of elasticity and lay the foundation for the evaluation of micro-macroscopic mechanical parameters such as Young's modulus ($E^*$), the Poisson ratio ($\nu^*$), and so on. Nothing to mention that parameters $E$ and $E^*$ as well as $\nu$ and $\nu^*$ are not the same so that their coincidence is quite accidental. Obviously, atomistic approach falls in opinion comparing with the continuum one due to time consuming calculations and, as a result, due to applicability to smaller objects. However, it possesses doubtless advantages concerning the description of the mechanical behavior of the object under certain loading (shape changing) as well as exhibiting the deformation and failure process at atomic level.

Recently a new atomistic approach has been suggested for the description of the graphene deformation based on considering the failure and rupture process of graphene as the occurrence of a mechanochemical reaction [29-32]. A similarity between mechanically induced reaction and the first-type chemical ones, first pointed out by Tobolski and Eyring more than sixty years ago [33], suggested the use of a well developed quantum-chemical approach of the reaction coordinate [34] in the study of atomic structure transformation under deformation. Firstly applied to the deformation of poly(dimethylsiloxane) oligomers [35], the approach has revealed a high efficacy in disclosing the mechanism of failure and rupture of the considered polymers. It has been successfully applied recently for the description of the uniaxial tension of both graphene [29, 30] and graphane [31] molecules, thus exhibiting itself as a significant part of the current molecular theory of graphene [28].

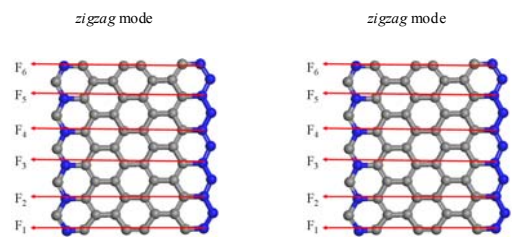

**Fig. 1**. Six mechanochemical internal coordinates of uniaxial tension of the molecule (5,5) NGr for two deformation modes. $F_1$, $F_2$, $F_3$, $F_4$, $F_5$ и $F_6$ are forces of response along these coordinates. Blue atoms fix the coordinates ends.

The main point of the approach concerns the reaction coordinate definition. When dealing with chemical reactions, the coordinate is usually selected among the internal ones (valence bond, bond angle or torsion angle) or is presented as a linear combination of the latter. Similarly, mechanochemical internal coordinates (MICs) are introduced as modified internal coordinates defined in such a way as to be able to specify the considered deformational modes [35, 37]. Thus, uniaxial tension and contraction are described by linear MICs similar to valence bonds. In the case of tensile deformation, the benzenoid pattern of graphene sheets and a regular packing of the units predetermined the choice of either parallel or normal MICs orientation with respect to the chain of C-C bonds. In the rectangular nanographene sheets and nanoribbons the former orientation corresponds to tensile deformation applied to the zigzag edges (*zigzag* mode) while the latter should be attributed to the armchair edges (*armchair* mode). The MIC configurations of the two tensile modes of the (5,5) NGr molecule are presented in Fig.1. The molecule has been led into the foundation of previously performed computational experiments [29-31] and presents a rectangular fragment of a graphene sheet cut along zigzag and armchair edges and containing 5 benzenoid units along each direction. The deformation proceeds as a stepwise elongation of the MICs with the increment $\delta L$=0.1Å at each step so that the current MIC length constitutes $L = L_0 + n\delta L$, where $L_0$ is the initial length of the MIC and $n$ counts the number of the deformation steps. Right ends of all the MICs are fixed so that these blue colored atoms are immobilized while atoms on the left ends of MICs move along the arrows providing the MIC successive elongation, once excluded from the optimization as well. The relevant force of response is calculated as the energy gradient along the MIC, while the atomic configuration is optimized over all of the other coordinates under the MIC constant-pitch elongation. The results presented in paper were obtained in the framework of the Hartree-Fock unrestricted (UHF) version of the DYQUAMECH codes [37] exploiting advanced semiempirical QCh methods (PM3 version [37]).

The corresponding forces of response $F_i$ applied along the $i^{th}$ MICs are the first derivatives of the total energy $E(R)$ over the Cartesian coordinates [35]:

$$\frac{dE}{dR} = \langle \varphi | \frac{\partial H}{\partial R} | \varphi \rangle + 2 \langle \frac{\partial \varphi}{\partial R} | H | \varphi \rangle + 2 \langle \frac{\partial \varphi}{\partial P} | H | \varphi \rangle \frac{dP}{dR} \quad (1)$$

Here, $\varphi$ is the wave function of atom of the ground state at fixed nucleus positions, $H$ presents the adiabatic electron Hamiltonian, and $P$ is the nucleus momentum. When the force calculation is completed, the gradients are re-determined in the system of internal coordinates in order to proceed further in seeking the total energy minimum by atomic structure optimization.

Forces $F_i$ are used afterwards for determining all required micro-macroscopic mechanical characteristics, which are relevant to uniaxial tension, such as the total force of response $F = \sum_i F_i$, stress $\sigma = F/S = \left(\sum_i F_i\right)/S$, where $S$ is the loading area, the Young's modulus $E = \sigma/\varepsilon$, where both stress $\sigma$ and the strain $\varepsilon$ are determined within the elastic region of deformation.

## 3. Computational results

Thus arranged computations have revealed that a high stiffness of the graphene body is provided by the stiffness of benzenoid units. The anisotropy of the unit mechanical behavior in combination with different packing of the units either normally or parallel to the body C-C bond chains lays the ground for the structure-sensitive mechanism of the mechanical behavior of the object that drastically depends on the deformation modes [29-31]. The elastic region of tensile deformation of both graphene and graphane (5, 5) NGr molecules is extremely narrow and corresponds to a few first steps of the deformation. The deformation as a whole is predominantly plastic and dependent on many parameters. Among the latter, the most important is chemical state of the molecule edge atoms [32].

Equilibrium structures of the molecule before and after uniaxial tension, which was terminated by the rupture of the last C-C bond coupling two fragments of the molecule, are shown in Fig. 2. Looking at the picture, two main peculiarities of the molecule deformation should be notified. First concerns the anisotropy of the deformation with respect to two deformational modes. Second exhibits a strong dependence of the deformation on the chemical composition of the molecule edge atoms. As mentioned above, the deformation anisotropy of graphene has been attributed to mechanical anisotropy of the constituent benzenoid units [29, 30]. The dependence of the deformation on the chemical modification of framing edge atoms has been revealed for the first time.

As seen in Fig. 2, when the edge atoms are bare and not terminated by other chemicals, the deformation behavior is the most complex. The mechanical behavior of the (5, 5) NGr molecule is similar to that of a tricotage sheet when either the sheet rupture has both commenced and completed by the rupture of a single stitch row (*armchair* mode) or the rupture of one stitch is 'tugging at thread' the other stitches that are replaced by still elongated one-atom chain of carbon atoms (*zigzag* mode). In the **Fig.**

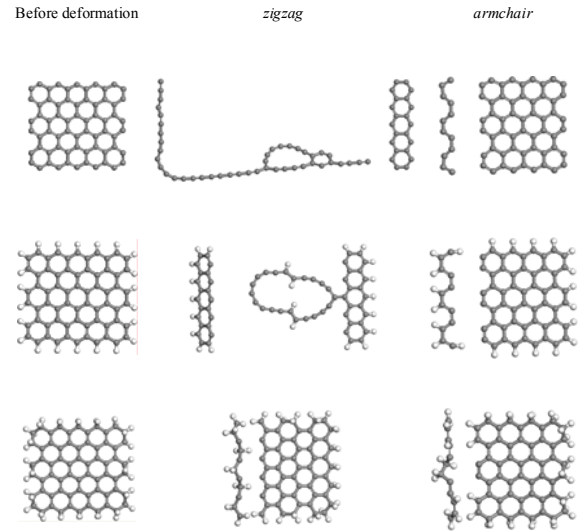

**2**. Equilibrium structures of the (5,5) NGr with different chemical modification of edge atoms before and after completing tensile deformation in two modes of deformation. Bare edges (top); $H_1$-terminated edges (middle); $H_2$-terminated edges (bottom).

former case, the deformation is one-stage and is terminated on the $17^{th}$ step of the deformation. In contrast, the deformational mode *zigzag* is multi-stage and consists of 250 consequent steps with elongation of 0.1Å at each step [29, 30]. The formation of one-atom chain under zizgzag-mode tension of a graphene sheet has been supported experimentally [39].

Quite unexpectedly, the character of the deformation has occurred to be strongly dependent on chemical situation at the molecule edges [31]. As seen in Fig. 2b, the addition of one hydrogen atom to each of the molecule edge atoms does not change the general character of the deformation: it remains a tricotage-like one so that there is still a large difference between the behavior of *zigzag* and *armchair* modes. At the same time, the number of the deformation steps of *zigzag* mode reduces to 125.

Even more drastic changes for this mode are caused by the addition of the second hydrogen atoms to the edge ones (Fig. 2c). Still, the *armchair* mode is quite conservative while *zigzag* one becomes practically identical to the former. The tricotage-like character of the deformation is completely lost and the rupture occurs at the $20^{th}$ step.

Figure 3 presents a set of 'stress-strain' relations that fairly well highlight the difference in the mechanical behavior of all the three molecules. Table 1 presents Young's modules that were defined in the region of the elastic deformation. As seen from the table, the Young's modules depend on the character of edge atom chemical modification. As shown in [31], elastic properties of big molecules such as polymers [35, 40] and nanographenes [31] are determined by dynamic characteristics of the objects, namely, by force constants of the related vibrations. Since benzenoid units have a determining resistance to any deformation of graphene molecules, the dynamic parameters of stretching C-C vibrations of the units are mainly responsible in the case of uniaxial tension. Changing in Young's modules means changing in the force constants (and, consequently, frequencies) of these vibrations. The latter are attributed to G-band of graphene that lays the foundation of a mandatory testing of any graphenium system by Raman spectrum. In many cases, the relevant band is quite wide which might indicate the chemical modification of the edge zone of the graphene objects under investigation.

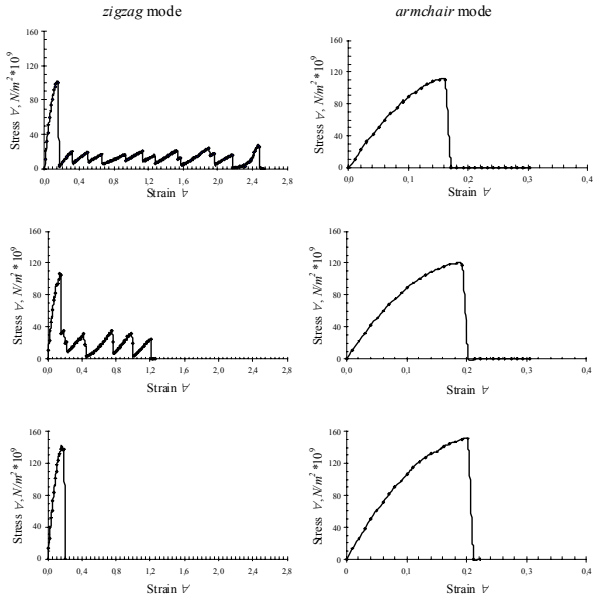

**Fig. 3.** Stress-versus-strain dependences of tensile deformation of the (5, 5) NGr molecule with different chemical modification in two deformation modes. Bare edges (top); $H_1$-terminated edges (middle); $H_2$-terminated edges (bottom).

Since the molecule deformation is mainly provided by basal atoms, so drastic changes in the deformation behavior points to a significant influence of chemical state of edge atoms on the electronic properties in the basal plane. The observed phenomenon can be understood if suggest that 1) the deformation and rupture of the molecule is a collective event that involves the electron system of the molecule as a whole; 2) the electron system of the graphene molecule is highly delocalized due to extreme correlation of odd electrons; and 3) the electrons correlation is topologically sensitive due to which chemical termination of edge atoms so strongly influences the behavior of the entire molecule. The latter has turned out to be a reality, indeed.

**Table 1.** Young's modules for (5,5) NGr with different configuration of edge atoms, TPa

| Mode | Bare edges | $H_1$-terminated edges | $H_2$-terminated edges |
|---|---|---|---|
| 'zigzag' | 1.05 | 1.09 | 0.92 |
| 'armchair' | 1.06 | 1.15 | 0.95 |

## 4. Topological character of the odd electron correlation in graphene

The performed computations have revealed that the correlation of the odd electrons of the studied molecules changes quite remarkably in due course of the deformation. This result can be illustrated by the evolution of the total number of effectively unpaired electrons $N_D$ during the deformation. The $N_D$ value is a direct characteristic of the extent of the electron correlation, on one hand, [2] and molecular chemical susceptibility, on the other, [41]. Changing in $N_D$ reveals changing in the molecule chemical activity induced by deformation.

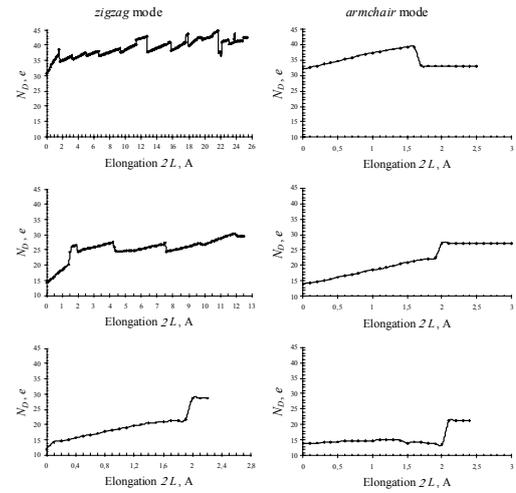

**Fig. 4.** Evolution of the odd electrons correlation interms of the total numbers of effectively unpaired electrons under tensile deformation of the (5, 5) NGr at two deformation modes. Bare edges (top); $H_1$-terminated edges (middle); $H_2$-terminated edges (bottom).

Figure 4 presents the evolution of $N_D$ for the three studied molecules. Since breaking of each C-C bonds causes an abrupt changing in $N_D$, a toothed character of the relevant dependences related to *zigzag* mode of the molecule with bare and H-terminated edges is quite evident. One should draw attention to the $N_D$ absolute values as well as to their dependence on both chemical modification of the edge atoms and the deformational modes. Evidently, chemical activity of the molecules is drastically changed in due course of a mechanically induced transformation, This changing is provided by the redistribution of C-C bond lengths caused by the mechanical action. This action combines positions of both basal plane and edge atoms into united whole and is topologically sensitive. Therefore, the redistribution of the C-C bonds over their lengths causes changing in a topological 'quality' of individual bonds. To illustrate the latter, let us look at not the total number of effectively unpaired electrons $N_D$, but at the atomic chemical susceptibility distribution over the molecule atoms that is determine by a fractional number of the effectively unpaired electrons $N_{DA}$ at atom $A$ [28].

### 4.1. Graphene molecule with bare edges

Figure 5 presents a set of the $N_{DA}$ image maps related to the first 17 steps of the armchair mode of the molecule uniaxial tension. The set corresponds to gradually increased $N_{DA}$ values up to the 17th set shown at the right-hand top panel of Fig.4. The maps are accompanied by the molecule equilibrium structures. For the maps to be presented in one scale, the $N_{DA}$ data were normalized by equalizing maximum $N_{DA}$ values at each map to that one at the zeroth step that corresponds to the non-deformed molecule. White figures present the equalizing coefficients.

As seen in the figure, the segregation of atoms into two groups related to edge and basal plane areas, respectively, which is characteristic for the non-deformed molecules, takes place up to the 16th step inclusively. Obviously, since simultaneously, the total number of the effectively unpaired electrons $N_D$ grows, a redistribution of the $N_{DA}$ values should occurs. However, the latter mainly concerns the basal plane atoms leaving the edge atoms only slightly changed due to practically constant maximum $N_{DA}$ values of the latter, which follows from the presented equalizing coefficient. Actually, Fig. 6 shows the distribution of the absolute $N_{DA}$ values for the zeroth and 16th steps alongside with their ratio. As seen in the figure, the redistribution concerns basal plane atoms mainly, indeed.

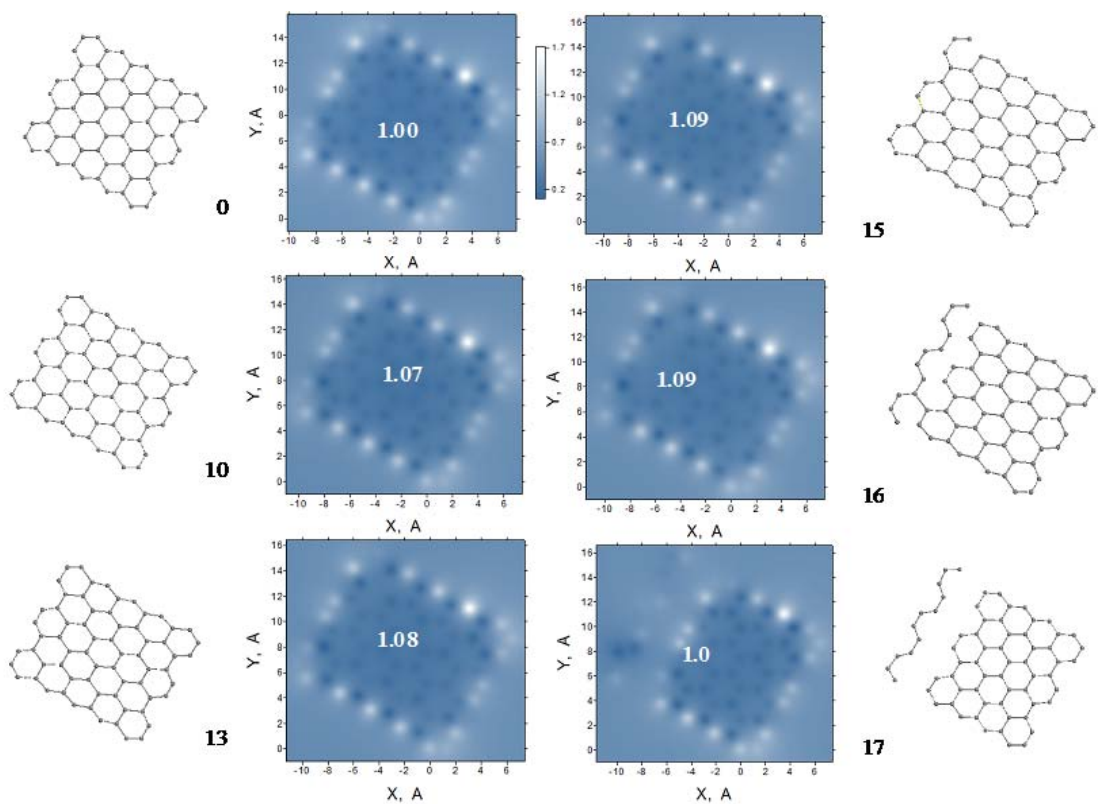

**Fig. 5.** Image $N_{DA}$ maps and equilibrium structures of the bare-edge (5, 5) NGr molecule in due course of the first stage armchair-mode tensile deformation. Black and white figures number steps and equalizing coefficients, respectively. Scale is related to all the maps.

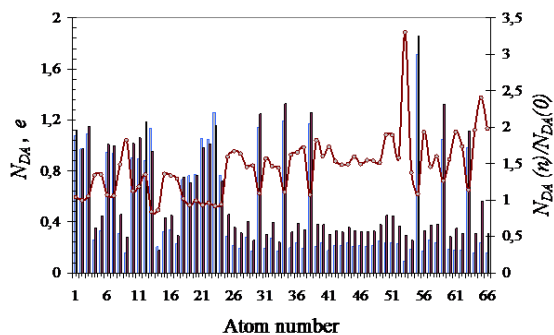

**Fig. 6.** Histograms of the $N_{DA}$ distribution over atoms of the bare-edge (5, 5) NGr molecule related to non-deformed (light) and 16th step-deformed (dark) armchair-mode tensile deformation. Curve plots the presented values ratio when $n=16$.

Important to draw attention to both image $N_{DA}$ maps and the corresponding structures related to the 15th and 16th steps. The structures are drawn by a visualization standard program that is based on tabulated values of chemical bond lengths. According to the data, there is one broken C-C bond at the 15th step and five of them at the 16th step. Since the bonds are ruptured homolytically, one should expect a radicalization of the molecule at these points. Providing the $N_{DA}$ indicative ability of just the very events, the appearance of white spots on the maps should be expected. However, until the 17th step no such spots are observed. This shows that the bonds breaking occurs at much longer interatomic distance in comparison with standard tabulated data, which supports a conclusion made in [42].

Figure 7 presents a similar picture describing the behavior of the odd electron correlation during zigzag-mode tension of the graphene molecule. The first 17 steps cover the first-stage zigzag-mode deformation and correspond to the first tooth of the strain-stress and $N_D$-elongation dependences shown in the left-hand top panels in Figs. 3 and 4, respectively. In contrast to the previous case, the redistribution of the values occurs quite differently. Although the general image of the maps is kept up to the 16th step, the appearance of strongly stretched C-C bonds can be already noticed at the 15th step while a complete bond breaking becomes absolutely evident at the 17th step only. The effect of stretched C-C bonds in the basal plane on the $N_{DA}$ value redistribution is well seen in Fig. 8. About 5.5-fold changing in the $N_{DA}$ value is characteristic for one of them while edge atoms keep only slightly changed $N_{DA}$ values.

As was mentioned previously [29, 30], the perstep elongation (stretching) of C-C bonds under *armchair*- and *zigzag*-mode tension is quite different from a geometrical viewpoint. Evidently, this might explain the difference in the $N_{DA}$ values related to the same step but cannot evidence why the character of the redistribution for the two modes is completely different thus attributing the difference to a topological nature of the considered mechanochemical reaction.

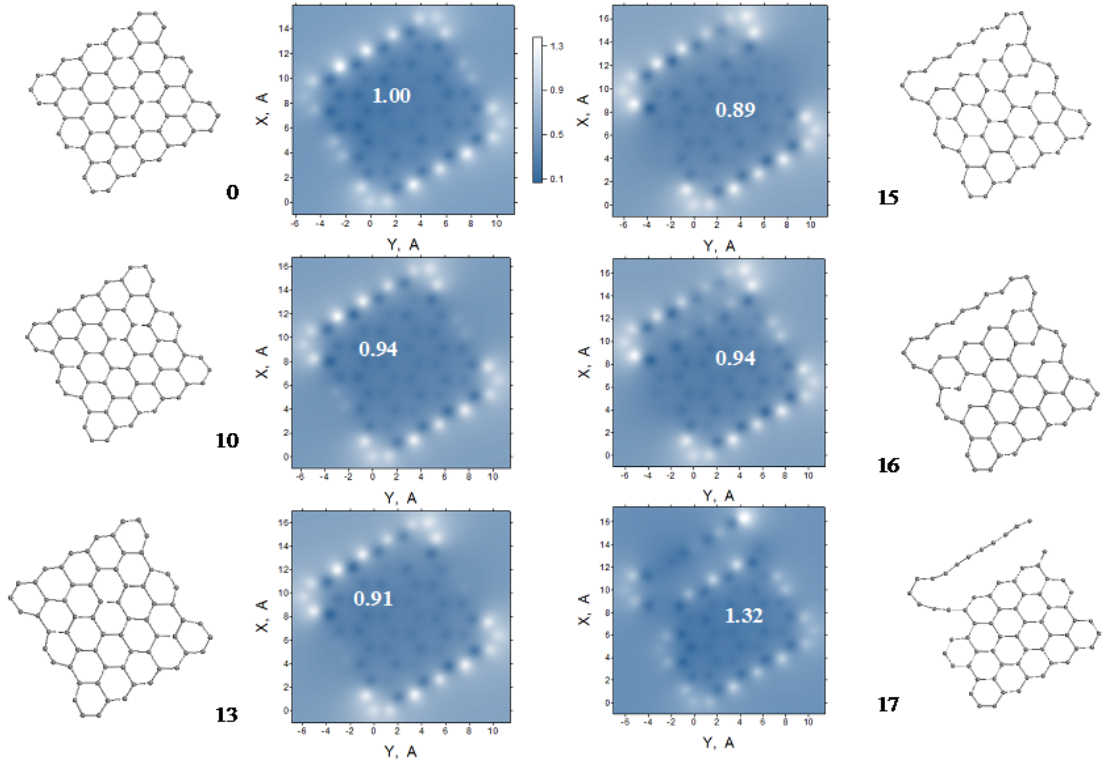

**Fig. 7.** Image $N_{DA}$ maps and equilibrium structures of the bare-edge (5, 5) NGr molecule in due course of the first stage zigzag-mode tensile deformation. Black and white figures number steps and equalizing coefficients, respectively. Scale is related to all the maps.

### 4.2. Single-hydrogen terminated edges of graphene molecule

The evolution of the $N_{DA}$ image maps in due course of the armchair- and zigzag-mode uniaxial tension of the (5, 5) NGr molecule with single-hydrogen terminated edges is presented in Fig. 9. The data are related to the first stage of deformation for both modes. The picture in the figure drastically differs from that one shown in Figs. 5 and 7. The difference starts from the non-deformed molecule. As seen in the figure, the $N_{DA}$ distributions differ remarkably for the two modes. The two maps are obtained in due course of the following procedure. Firstly, the equilibrium structure of the free standing $H_1$-terminated (5, 5) NGr molecule was obtained. Afterwards, a set of carbon atoms which fix six selected MICs (see blue atoms in Fig.1) was fixed, differently for two deformation modes. Then the optimization procedure was repeated for adopting calculation to the new conditions. If in the case of the bare edge molecule considered in the previous section, this procedure caused only a homogeneous scaling of the $N_{DA}$ values (see scales in Figs. 5 and 7), a complete reconstruction of the image $N_{DA}$ map occurs in the current case while the maximum $N_{DA}$ values are practically identical. Both findings mean that fixation of the graphene sheet edges provides a considerable change in its electron distribution. The latter is additionally greatly influenced by the chemical composition of the sheet edge atoms.

Coming back to Fig.9, one can see that, as previously, in the case of *armchair* mode the image maps keep practically unchanged appearance until the 19[th] step in spite of highly stretched C-C bonds at the latter step. The next 0.1Å elongation provides simultaneous breaking of six C-C bonds followed with 3-fold increasing of the $N_{DA}$ values. The sample becomes highly radicalized. Concentration of high $N_{DA}$ values in the area of broken bonds drastically changes the image map fully suppressing much less active pristine atoms. Further elongation does not change the situation.

In the case of *zigzag* mode, the map appearance has been kept up to the 14[th] step after which highly stretched C-C bonds are observed in the middle of the molecule basal plane at the 15[th] step, whose complete breaking is followed at the 16[th] step. The sample becomes highly radicalized.

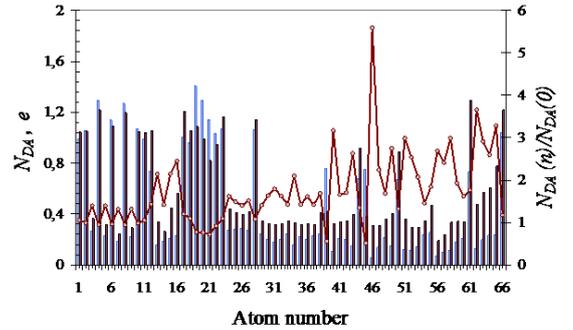

**Fig. 8.** Histograms of the $N_{DA}$ distribution over atoms of the bare-edge (5, 5) NGr molecule related to non-deformed (light) and 16[th] step-deformed (dark) zigzag-mode tensile deformation. Curve plots the presented values ratio when $n=16$.

### 4.2. Double-hydrogen terminated graphene molecule

Addition of the second hydrogen atom to the edge ones drastically changes the image maps again as seen in Fig. 10. In contrast to the previous case, the initial fixation of edge atoms does not cause changing in both the map appearance and absolute $N_{DA}$ value. However, the molecule becomes non-planar, which greatly influence further deformation. Thus, in due course of the armchair-mode tension, the difference in the values of eight framing basal atoms and remainders is gradually smoothed once equalizing at the 19[th] step. The situation remains the same for the 20[th] step in spite if the presence of stretched C-C bonds. The bond breaking occurs at the 21$^{st}$ step, the sample becomes radicalized with a small area of the radical concentration. Oppositely to the case, the zigzag-mode deformation does not cause any smoothing of the $N_{DA}$ values distribution and keeps the non-deformed shape up to the 19$^{th}$ step. The bond stretching is observed at steps from 17$^{th}$ to 19$^{th}$ and the bond breaking occurs at the 20$^{th}$ step.

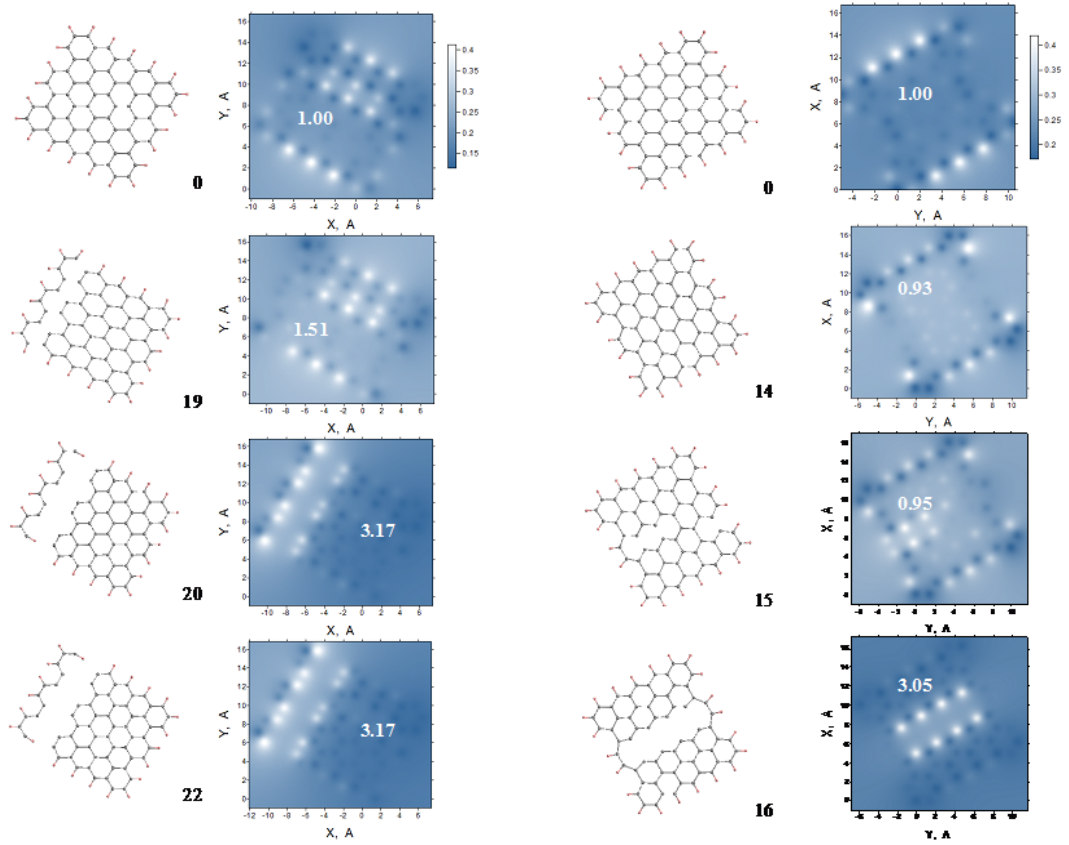

**Fig. 9.** Image $N_{DA}$ maps and equilibrium structures of the $H_1$-terminated-edge (5, 5) NGr molecule in due course of the first stage of the armchair (left) and zigzag (right) -mode tensile deformation. Black and white figures number steps and equalizing coefficients, respectively. Scales are related to all the maps within the deformational mode.

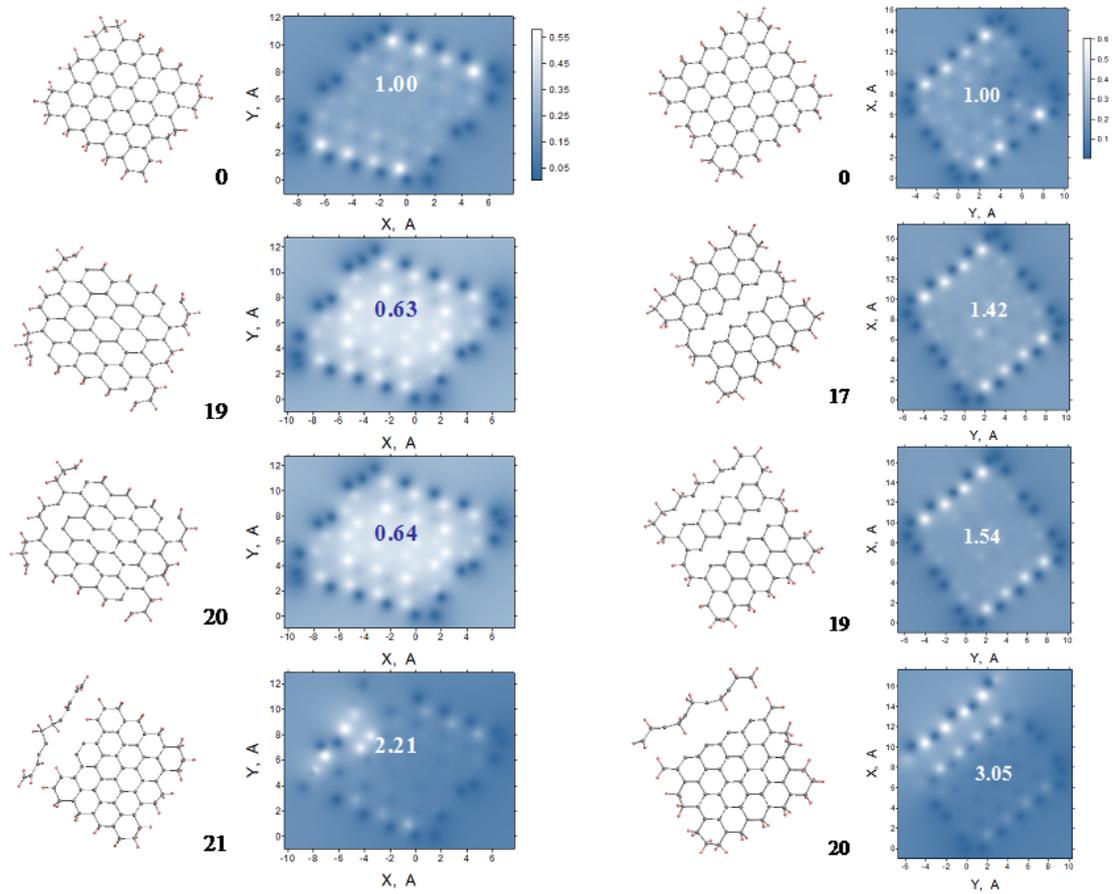

**Fig. 10.** Image $N_{DA}$ maps and equilibrium structures of the $H_2$-terminated-edge (5, 5) NGr molecule in due course of the first stage of the armchair (left) and zigzag (right) -mode tensile deformation. Black and white figures number steps and equalizing coefficients, respectively. Scales are related to all the maps within the deformational mode

Taking together, Figs. 5, 7, 9, and 10 exhibit changing odd electron correlation of the graphene molecule under deformation and highlight a strong dependence of the correlation on both the deformational mode configuration and chemical modification of the molecule edge atoms.

## 5. Conclusion

Presented in the current paper undoubtedly shows that the chemical modification of the graphene molecule edge atoms has a great impact on its mechanical behavior. The feature is a result of a significant correlation of the molecule odd electrons followed by their conjugation over the molecule. Thus, the transition from the molecule with bare edges, characterized by maximal correlation of odd electrons, to the molecule with $H_1$- and $H_2$-terminated edges is followed by a considerable suppression of the correlation related to the edge atoms in the former case and a complete zeroing of the latter in the second case. As turned out, the changes are not local and strongly influence the electronic structure in the region of the basal plane, where the main deformational process occurs, causing a redistribution of C-C bonds over their lengths, thus, changing 'the quality' of the bonds and providing a topological character of deformational processes in graphene.